\documentclass[twocolumn, showpacs, amsmath,amsfonts, superscriptaddress,prb]{revtex4-1}
\usepackage{graphicx}
\usepackage{amsmath}
  \usepackage{amssymb}
  \usepackage{bm}

\usepackage[cp1251]{inputenc}
\usepackage[T1,T2A]{fontenc}

\usepackage{color}

\begin{document}
\title{Intense Anti-Stokes Emission of Erbium Ions in Gallium Lanthanum Sulphide-Oxide Glass
in Visible Spectral Range}

\author{
  M. M. Voronov$^{\ast}$\textsuperscript{\textsf{\bfseries 1}},
  \email{mikle.voronov@coherent.ioffe.ru}
  A. B. Pevtsov\textsuperscript{\textsf{\bfseries 1}},
  A. P. Skvortsov\textsuperscript{\textsf{\bfseries 1}},
  C. Koughia\textsuperscript{\textsf{\bfseries 2}},
  C. Craig\textsuperscript{\textsf{\bfseries 3}},
  D. W. Hewak\textsuperscript{\textsf{\bfseries 3}},\\
  S. Kasap\textsuperscript{\textsf{\bfseries 2}},
   V. G.    Golubev\textsuperscript{\textsf{\bfseries 1}}.}

 \affiliation{$^1$Ioffe  Institute, Politekhnicheskaya 26, St Petersburg, 194021 Russia\\
 $^2$Department of Electrical and Computer Engineering, University of Saskatchewan, Saskatoon, Canada\\
 $^3$Optoelectronics Research Centre, University of Southampton, Southampton, UK SO17 1BJ}

\begin{abstract}
Photoluminescence spectra have been investigated in Er$^{3+}$ doped GaLaS(O) glasses. The samples
demonstrate intense  ``green'' emission bands centered at around 525 and 550 nm due to up-conversion
processes in Er$^{3+}$ ions. The theoretical description of up-conversion intensity as a function of
excitation intensity has been offered. It is based on a solution of a system of rate equations taking into
account three up-conversion transitions.
\end{abstract}

 \maketitle

\section{Introduction}
Rare-earth doped chalcogenide glasses demonstrating effective up-conversion are   attractive for many
practical applications including up-conversion fiber optical lasers \cite{Scheps}, up-converters for solar
cells to convert Sun's IR radiation into visible \cite{Kharel} and  others. These glasses  may  also  be
used as doped glasses where, due to their exceptionally low phonon energy, the up-conversion effects may
investigated at its fullest.

 In the present paper we investigate the intense visible green and red up-conversion in gallium lanthanum
 sulphide-oxide glass GaLaS(O) doped with trivalent Er$^{3+}$  erbium ions \cite{Ravagli}. These glasses may
 demonstrate high efficiency up-conversion as well as high quantum yield
 of ``regular'' photoluminescence
 (PL) presumably due to insignificant non-radiative losses which are typical for these glasses.
 The addition of oxygen into the well studied glass matrix GaLaS \cite{Hewak} shifts the optical
 absorption edge towards higher energies further improving  up-conversion conditions by suppressing
 losses due to energy exchange between electron subsystem of  chalcogenide matrix and excited levels
 of the erbium ion \cite{Tverjanovich}.

Luminescent properties of ions with discreet sets of energy levels may be reasonably well described
theoretically by a set of rate equations \cite{Carroll} taking into account several lower energy levels.
In reality, these levels may be complicated manifolds consisting of overlapping Stark
levels broadened by the presence of a non-crystalline matrix. In the present paper, we offer a simple
phenomenological method to calculate the PL intensity by using the appropriate functions.
We stick to ten energy level ionic system which has been proven to be good for the trivalent erbium ion.
Under external laser induced excitation, this system shows four upward transitions: one
transition from the ground state and three transitions using excited state absorption (ESA).
The method, which is  developed in the  present paper, is used to describe the experimentally observed spectra of
up-conversion PL in Er$^{3+}$ ions embedded in a gallium lanthanum sulphide-oxide glass.

\section{Experimental details}
The Er$^{3+}$ doped gallium lanthanum sulphide-oxide glass was prepared by melt quenching of raw materials
mixed in the following proportions: 72.5\%  Ga$_2$S$_3$, 27\% La$_2$O$_3$ and doped with  0.5\%  Er$_
2$O$_3$. All percent are molar. Melting took place in a vitreous carbon crucible for 24 hours in dry
flowing argon followed by annealing at 500$^{\circ}$C, ramping up and down at  1$^{\circ}$ per minute.
\begin{figure}[t]
\includegraphics*[width=0.9\linewidth,height=\linewidth]{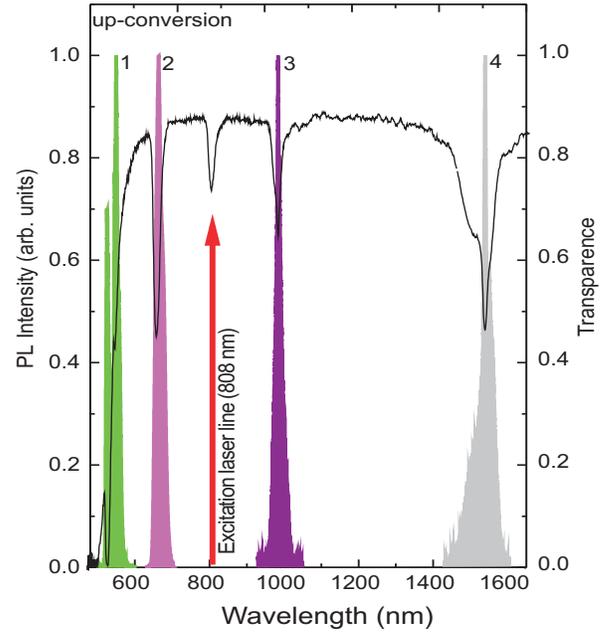}
\caption{
  \label{Fig.1}Comparison of photoluminescence spectrum of Er$^{3+}$ doped GaLaS(O) glass (left scale)
  with optical transmittance (solid line, right
scale). Emission bands correspond to ($^2$H$_{11/2}$, $^4$S$_{3/2}$) --$^4$I$_{15/2}$ (1), $^4$F$_{9/2}$
--$^4$I$_{15/2}$ (2), $^4$I$_{11/2}$ --$^4$I$_{15/2}$ (3), $^4$I$_{13/2}$ --$^4$I$_{15/2}$ (4)
transitions. Excitation by a 808 nm laser diode.}
 \label{onecolumnfigure}
\end{figure}

The PL was excited by laser diode operating at a wavelength of 808 nm which coincides with the
$^4$I$_{15/2}$ --$^4$I$_{9/2}$ absorption band of Er$^{3+}$. Glass samples doped with Er$^{3+}$ showed
bright green emission well seen by naked eye under regular ambient daylight illumination.
 The transmittance and PL spectra in the 400-1700 nm range  were measured using a 600 mm single-grating
spectrometer equipped with Si and InGaAs CCDs. The spectra were taken at room temperature and corrected
for the system response.

Figure 1 gives an overview of the experimental results. Under 808 nm excitation several PL bands with
Stokes and anti-Stokes shifts have been observed. Two bands centered at 980 nm and 1535 nm have
demonstrated ``normal'' Stokes shift and are easily assigned to $^4$I$_{11/2}$ --$^4$I$_{15/2}$ and
$^4$I$_{13/2}$ --$^4$I$_{15/2}$ transitions, respectively. The bands with anti-Stokes shift show a ``red''
peak centered at around 664 nm due to, presumably, $^4$F$_{9/2}$ --$^4$I$_{15/2}$ transition and a
``green'' band consisting of two peaks centered at 525 and 550 nm which may be related to $^2$H$_{11/2}$
--$^4$I$_{15/2}$ and $^4$S$_{3/2}$ --$^4$I$_{15/2}$ transitions, respectively.

\begin{figure}[t]
\includegraphics*[width=\linewidth,height=\linewidth]{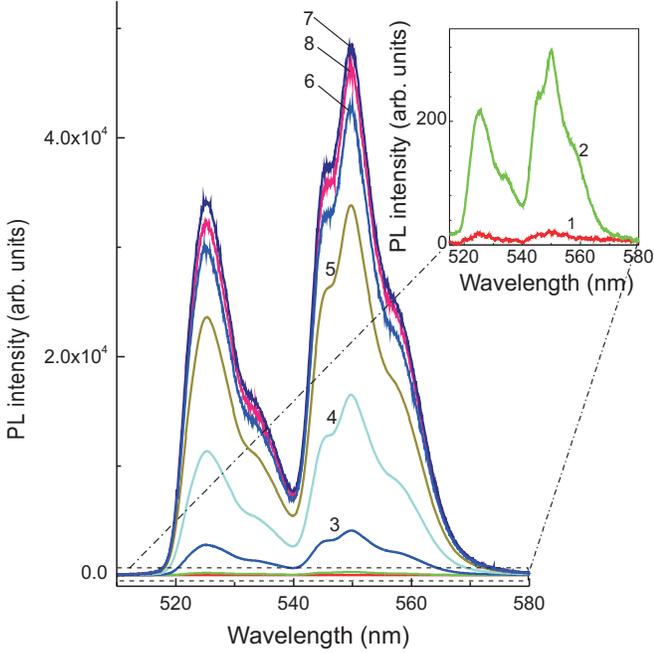}
\caption{
  \label{Fig2}  Emission spectra corresponding to $^2$H$_{11/2}$ --$^4I_{15/2}$ and $^4$S$_{3/2}$
--$^4$I$_{15/2}$ transitions at varying excitation intensities: 3 W/cm$^2$ (1), 30 W/cm$^2$ (2), 300
W/cm$^2$ (3), 750 W/cm$^2$ (4), 1700 W/cm$^2$ (5), 4200 W/cm$^2$ (6), 8500 W/cm$^2$ (7), 17000 W/cm$^2$
(8). Excitation by a 808 nm laser diode.}
 \label{onecolumnfigure}
\end{figure}

Figure 2 demonstrates the influence of excitation intensity ($P$) on the line shape and intensity of the
``green'' line. It is worth noting  that the line  shape remains virtually unchanged as well as the ratio
of peak intensities of emission bands corresponding to $^2$H$_{11/2}$ --$^4$I$_{15/2}$ and $^4$S$_{3/2}$
--$^4$I$_{15/2}$ transitions as confirmed by Figure 3. Figure 3 also shows that the intensities of both
above mentioned bands experience monotonous increase with the increase of excitation intensity up to
$P\approx$ 2000~W/cm$^2$  followed by eventual saturation.

  \section{Theoretical analysis and discussion}
  Figure 4 shows the schematic diagram of ten discreet level model used to explain experimental data.
  One of the key question of this approach (which is usually  omitted) is the validity
  of presentation of involved Er$^{3+}$ ion manifolds by discreet levels. This can be
  valid if ``the time required to establish a thermal distribution within each manifold
  is short compared with the lifetime of this manifold'' \cite{Miniscalco}. The later assumption may be
  easily verified by testing the validity of the McCumber relation between
  emission ($\sigma_e$) and absorption ($\sigma_a$) cross-sections which is derived assuming
 a thermal equilibrium within the manifold
\begin{equation*}
 \sigma_e=\sigma_a(\nu)\exp\left(\frac{\varepsilon-h\nu}{kT}\right),
  \end{equation*}
here $\varepsilon$  is a temperature dependent excitation energy, $h\nu$  is the photon energy
\cite{McCumber}. It is interesting to note that the above relation turns out to be valid not only for all
separated manifolds involved in our model but it holds also for a group of manifolds of $^4$S$_{3/2}$ and
$^2$H$_{11/2}$. Additionally, this approach is supported by the independence of line shape of  ``green''
band from excitation intensity (Figure 2) and, hence, the stability of the ratio of peak intensities
corresponding to $^2$H$_{11/2}$ --$^4$I$_{15/2}$ and $^4$S$_{3/2}$ --$^4$I$_{15/2}$ transitions as shown
in Figure 3.
\begin{figure}[t]
\includegraphics*[width=0.95\linewidth,height=\linewidth]{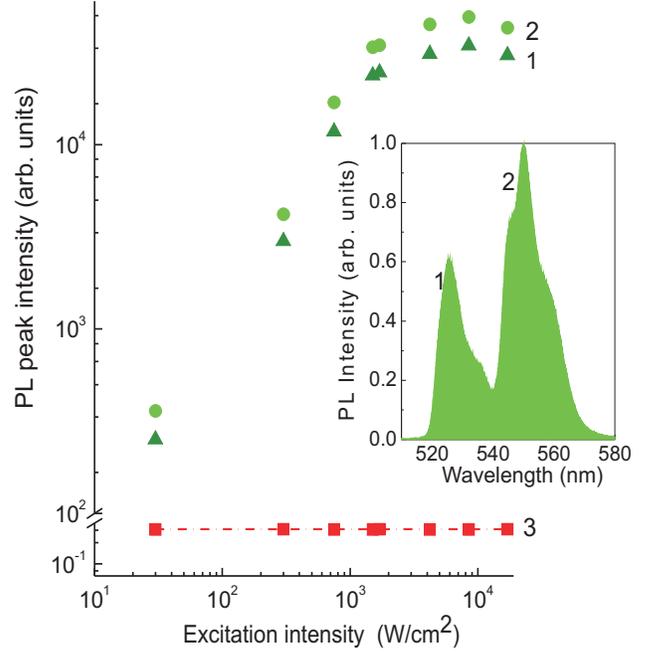}
\caption{
\label{Fig3}Peak intensities of emission spectra (see inset) corresponding to $^2$H$_{11/2}$
--$^4$I$_{15/2}$ (1) and $^4$S$_{3/2}$ --$^4$I$_{15/2}$  (2) transitions as a function of excitation
intensity. Red squares (3) is the ratio of (1) to (2) peak intensities. Broken line is a guide to the
eye.} \label{onecolumnfigure}
\end{figure}
The pumping is considered to be to the manifold $^4$I$_{9/2}$ corresponding to the 808 nm excitation used
in experiments. Figure 4 shows the existence of three possible up-conversion transitions (based on the ESA
mechanism) via excited states $^4$I$_{13/2}$, $^4$I$_{11/2}$ and $^4$I$_{9/2}$. It shows also a set of
radiative transitions part of which was observed experimentally as PL spectra (see Figure 1). Other long
wavelengths transitions fall outside of the sensitivity range of our experimental installation while the
transitions from $^2$H$_{9/2}$, $^4$F$_{3/2}$ and $^4$F$_{5/2}$ are disguised or/and suppressed by strong
matrix absorption. We could not observe the $^4$I$_{9/2}$ --$^4$I$_{15/2}$ emission band, which strongly
overlaps the excitation spectrum.
\begin{figure}[t]
\includegraphics*[width=\linewidth,height=\linewidth]{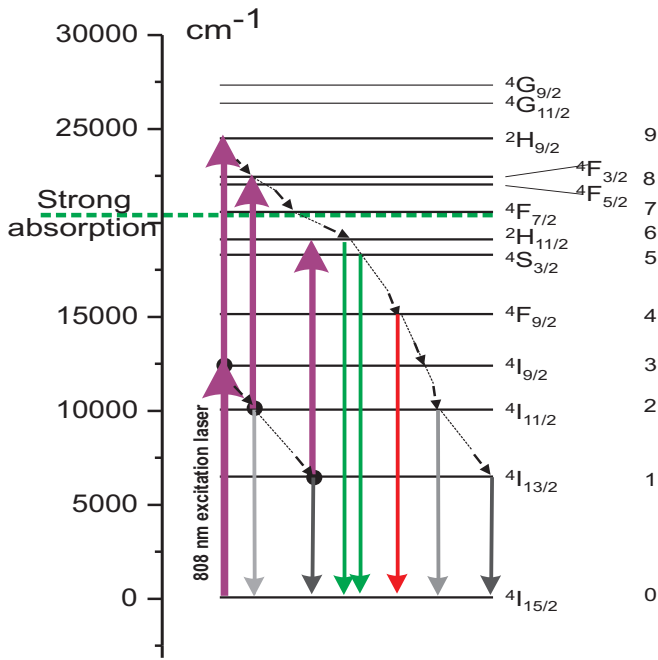}
\caption{ \label{Fig4}Schematic diagram of Er$^{3+}$ ion manifolds and interlevel transitions used in the
theoretical model. Diagram shows the optical transition from ground state (GSA), three up-conversion
transitions using ESA as well as experimentally observed emission transitions. Term ``strong absorption''
shows the energy at which absorption coefficient of matrix glass exceeds at least two times the strongest
absorption band of Er$^{3+}$.}
 \label{onecolumnfigure}
\end{figure}
In general, finding the relation between the intensity of up-converted PL and excitation intensity starts
with the creation of a system of rate equations describing the radiative and non-radiate transitions
between various energy levels. The solution of this system gives the populations of all levels involved
under consideration, which allows the calculation of the intensities of all spectral lines corresponding
to optical transitions.  Deeper analysis may involve the laborious decomposition of PL spectra into a set
of Gaussians corresponding to transitions between individual Stark levels (similar to what was done in
\cite{Koughia}. Each of the above Gaussians is in fact an envelope of individual Lorentzians (in
crystalline host material) or individual Gaussians (in non-crystalline host material). Even more detailed
calculations should include induced transitions and modification of emission spectra due to absorption and
spatial non-uniformity of specimen relative permittivity.

Schematic diagram of Er$^{3+}$ ion manifolds \cite{Gruber} and interlevel transitions used in the theoretical model.
Diagram shows the optical transition from ground state (GSA), three up-conversion transitions using ESA as
well as experimentally observed emission transitions. Term ``strong absorption''  shows the energy at
which absorption coefficient of matrix glass exceeds at least two times the strongest absorption band of
Er$^{3+}$.

In the present paper, we limit the model by calculating the intensities of  up-conversion PL bands as a
function of excitation intensity ($P$). We offer a theoretical description of experimental results by
analyzing a set of radiative and non-radiative transitions within nine excited and one ground level
simulating manifolds of Er$^{3+}$ ion (Figure 4). The corresponding system of rate equations along with
additional equation of conservation of total number of electrons may be presented in a matrix form as
$\hat{\Gamma}\bf{n}=\bf{b}$ where the vectors $\bf{n}$ and $\bf{b}$ are defined as
${\bf{n}}=(n_0,n_1,...n_9)^{T}$, ${\bf{b}}=(1,0,0,...0)^{T}$. Matrix $\hat{\Gamma}$ has dimensions
$10\times10$, that allows us to solve the system of equations by using the Cramer rule. This approach
potentially allows us to calculate the coefficients included in the expressions for optical transition
intensities. However, in multilevel ionic systems, dozens or even hundreds of addends must be taken into
account, with every addend being a product of several parameters such as generation and relaxation rates.
This type of calculation is quite sensitive to uncertainties of above mentioned parameters and may lead to
considerable discrepancies. Instead, general equations for level populations may be represented by ratios
of polynomials (4th or 3rd orders), which may be simplified further by taking into account the small
numbers of addends at zeroth and first degrees of generation rate. As a result, the general equation for
radiative transition intensity \textit{J} may be reduced to
\begin{equation}
 J(I)=\frac{AI^2+BI+C}{DI^2+EI+F},
 \end{equation}
where \textit{I} is the pumping intensity and \textit{A}, \textit{B, C, D, E} and \textit{F} are numerical
coefficients. Moreover, for optical transitions involving  lower three levels, the coefficient \textit{A}
is simply 0, while for higher levels with a high accuracy \textit{C} is equal to 0.
 Figure 5 demonstrates the quality of this approximation for two up-conversion bands in GaLaS(O):Er
glass. The appropriate adjustable parameters are summarized in the caption to Figure 5.

The advantage of this approach is that one may directly use Eq. (1) to fit the interpolation curve through
well-established experimental points and to use optimized $A-F$ parameters for further predictions.
Therefore, it seems reasonable to establish some general features of function $J(I)$ which may be done by
calculating and zeroing first and second derivatives, and by finding the number of possible solutions,
which leads to
$$
BDI^2 + 2CDI + CE-BF=0
$$
\begin{eqnarray}
&& BD^2I^3 + 3CD^2I^2 + 3(CE-BF)DI+ \nonumber\\&&
(CE-BF)E - CDF = 0\:.\nonumber
\end{eqnarray}
Mathematically, the function $J(I)$ may be very ``versatile'': i) it may be monotonic, monotonically
reaching saturation; ii) it may have only one maximum; iii)
    it may have two extrema (local minimum and (local) maximum).
    Besides, it may have up to three inflection points or  have none.
\begin{figure}[t]
\includegraphics*[width=0.9\linewidth,height=9cm]{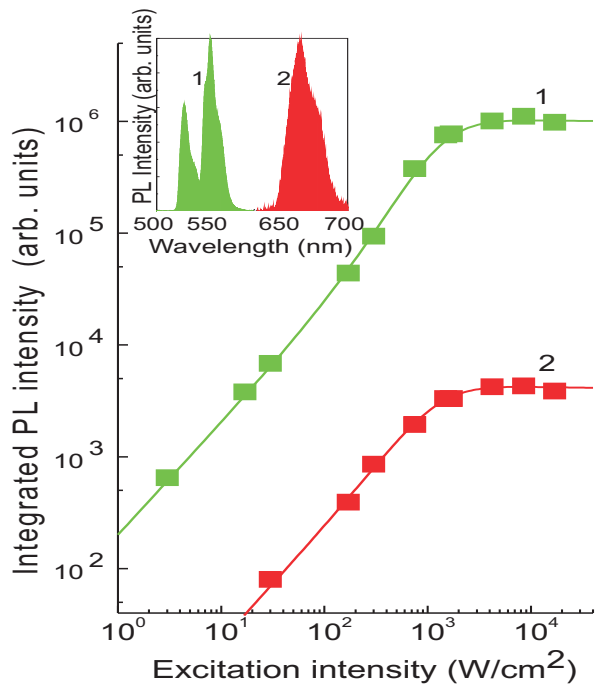}
\caption{ \label{Fig5}The integrated  intensities of ``green'' (1) and ``red ''  (2) photoluminescence
bands versus excitation intensity. Inset: Emission spectra corresponding to ($^2$H$_{11/2}$,
$^4$S$_{3/2}$) --$^4I_{15/2}$ and $^4$F$_{9/2}$ --$^4$I$_{15/2}$ transitions. Solid lines are theoretical
fits (Eq.1) with the following coefficients: $A=15530$,  $B=4116674$, $C=0$, $D=0.01487$, $E=1$,
$F=20920.4$ (curve 1); $A=38.94$, $B=49654.7$, $C=0$, $D=0.01039$, $E=1$, $F=23875.7$ (curve 2).}
 \label{onecolumnfigure}
\end{figure}
    For transitions from upper levels (marked 4, 5,...9 in Figure 4) $J(I)\rightarrow A/D$ at $I\rightarrow\infty$;
    for transitions from lower levels (marked 1, 2, 3 in Figure 4) $J(I)\rightarrow 0$ at $I\rightarrow\infty$.
Therefore, mathematically, there are several potentially possible scenarios while investigating PL in
these multilevel ionic systems. Physically, non-monotonic $J(I)$ behavior is connected with the existence
of several up-conversion transitions involving ESA. Those may also explain $J(I)$ saturation (assuming
 negligible induced emission) at $P$ exceeding 2000  W/cm$^2$ which is observed experimentally (Figure
5).

\section{Conclusion}
Intensive green red up-conversion has been detected in Er$^{3+}$ doped GaLaS(O) glass under laser diode
excitation at 808 nm. The dependence of up-conversion intensity as a function of pumping intensity has
been investigated. This dependence may be described by applying a phenomenological method developed in the
present paper. The method uses analytical expressions derived from rate equations taking into account
radiative and non-radiative interlevel transitions (including excited state absorption). The coefficients
entering these analytical expressions are defined by interpolation through several well-established
experimental points.

\vspace{1 mm}

\begin{acknowledgements}
Authors from Ioffe Institute thank the State assignment no. 0040-2019-0012  for financial support.
\end{acknowledgements}

\end{document}